\documentclass{article}
\usepackage{arxiv}
\usepackage{caption}
\usepackage[utf8]{inputenc} 
\usepackage[T1]{fontenc}    
\usepackage{hyperref}       
\usepackage{url}            
\usepackage{booktabs}       
\usepackage{amsfonts}       
\usepackage{nicefrac}       
\usepackage{microtype}      
\usepackage{lipsum}
\usepackage{graphicx}
\graphicspath{ {./images/} }

\usepackage{amsthm}
\usepackage{physics}
\usepackage{notoccite}
\usepackage{comment}

\newcommand{\VB}{V$_\text{B}^-$\space}
\newcommand{\mspm}{$\ket{m_\text{S}=\pm1}$\space}
\newcommand{\mszero}{$\ket{m_\text{S}=0}$\space}
\newcommand{\msminus}{$\ket{m_\text{S}=-1}$\space}
\title{Excited-state spin-resonance spectroscopy of V$_\text{B}^-$ defect centers in hexagonal boron nitride}

\author{
 Nikhil Mathur$^*$ \\
  School of Applied and Engineering Physics, Cornell University, Ithaca, NY, USA
   \And
 Arunabh Mukherjee$^*$  \\
  The Institute of Optics, University of Rochester, Rochester, NY, USA
  \And
 Xingyu Gao \\
  Department of Physics and Astronomy, Purdue University, West Lafayette, IN, USA
  \And
 Jialun Luo \\
  Department of Physics, Cornell University, Ithaca, NY, USA
  \And
  Brendan A. McCullian \\
  School of Applied and Engineering Physics, Cornell University, Ithaca, NY, USA
  \And
 Tongcang Li \\
  Department of Physics and Astronomy, Purdue University, West Lafayette, IN, USA\\
  Elmore Family School of Electrical and Computer Engineering\\ Purdue University, West Lafayette, IN 47907, USA
  \And
 A. Nick Vamivakas$^\dagger$ \\
 The Institute of Optics, University of Rochester, Rochester, NY, USA\\
 Materials Science, University of Rochester, Rochester, NY, USA\\
 Department of Physics and Astronomy, University of Rochester, Rochester, NY, USA\\
 Center for Coherence and Quantum Optics, University of Rochester, Rochester, NY, USA
  \And
 Gregory D. Fuchs$^{**}$ \\
  School of Applied and Engineering Physics, Cornell University, Ithaca, NY, USA\\
  Kavli Institute at Cornell for Nanoscale Science, Ithaca, NY, USA\\
  \And
  \\ *denotes equal contribution\\$^{\dagger}$nick.vamivakas@rochester.edu\\$^{**}$gdf9@cornell.edu  
}

\begin{document}
\maketitle
 \newpage
\begin{abstract}
The recently discovered spin-active boron vacancy (V$_\text{B}^-$) defect center in hexagonal boron nitride (hBN) has high contrast optically-detected magnetic resonance (ODMR) at room-temperature, with a spin-triplet ground-state that shows promise as a quantum sensor. Here we report temperature-dependent ODMR spectroscopy to probe spin within the orbital excited-state. Our experiments determine the excited-state spin Hamiltonian, including a room-temperature zero-field splitting of 2.1 GHz and a g-factor similar to that of the ground-state. We confirm that the resonance is associated with spin rotation in the excited-state using pulsed ODMR measurements, and we observe Zeeman-mediated level anti-crossings in both the orbital ground- and excited-state. Our observation of a single set of excited-state spin-triplet resonance from 10 to 300 K is consistent with an orbital-singlet, which has consequences for understanding the symmetry of this defect. Additionally, the excited-state ODMR has strong temperature dependence of both contrast and transverse anisotropy splitting, enabling promising avenues for quantum sensing. 
\end{abstract}


\section{Introduction}
Optically addressable, spin-active defects and quantum dots in the solid-state have emerged as promising qubits and quantum sensors\cite{Atature_review,Fuchs_NV_naturePhys,Togan2010} because robust control techniques enable facile quantum gates and sensing protocols \cite{Degen_RevModPhys2017}. The recent advent of two-dimensional (2D) materials has stimulated the search for spin-active defects that can be integrated into van der Waals heterostructures, enabling a wide array of optoelectronic and nanophotonic devices that take advantage of its optical and spin properties \cite{Geim2013,CCtrions2018,Ajit2019,Mukherjee2020,AIPMukherjee2020,ZhongScAdv2017,Zhong2020,PRRcrx3,Li2021,Froch_nanoLett2021}. A spin-active defect in a 2D material is especially promising for nanoscale sensing of interfacial phenomena with high sensitivity due to narrow spin transition linewidths and the ability to position these atomic-scale systems at sub-nanometer distances from the surface of a sample\cite{Tetienne_NatPhysNews,Gottscholl_sensing}.


 
\if In this context, optically accessible single spins have been studied in 2D semiconductors such as WSe$_2$, where optical transitions of localized charged excitons to spin-$\frac{1}{2}$ provide optical access to the spin-active ground states of single charge carriers at cryogenic temperatures\cite{CCtrions2018,Ajit2019,Mukherjee2020}.\fi Interestingly, the wide bandgap 2D material hexagonal boron nitride (hBN), which has been known to host bright and stable single-photon emitting defects\cite{Tran_firsthBN,Jungwirth_nanoLett,Jungwirth_PRL,Konthasinghe:19,Bassett_hBN,Proscia_nanopillar2018}, is now also known to host spin-active defects that are addressable at room temperature\cite{Gottscholl_natureMat,Chejanovsky2021,Xingyu_ACSPhoton2021,Tongcang_plasmon,Gottscholl_SciAdv2021}. Significant progress has been made in understanding the spin-active orbital ground-state (GS) of the negatively charged boron vacancy (V$_\text{B}^-$) defect, which is an orbital-singlet and spin-triplet with zero-field electron spin-resonance at 3.5 GHz arising from spin-spin interactions\cite{Gottscholl_natureMat}. However, there are only tentative proposals for the energy level structure of the excited-state (ES) as well as the overall symmetry of the defect, without experimental confirmation. The highest point group symmetry of the \VB defect is D$_{3h}$, with allowed optical transitions between the $^{3}A_{2}^{'}$ GS and the $^{3}E^{"}$ ES\cite{Gali2020,Reimers_PRB_2020}, but it is expected that symmetry breaking due to strain may result in a lowered C$_{2v}$ symmetry, splitting the ES orbital-doublet into non-degenerate orbital-singlet levels. In comparison to the nitrogen-vacancy (NV$^-$) center in diamond, where careful understanding of the ES Hamiltonian\cite{FuchsPRL2008,BatalovPRL2009,KaiMeiFu_PRL2009,Fuchs_natPhys_2010,FuchsPRL2012,Robledo_njp2013} was instrumental for key advances including spin readout enhancement\cite{Robledo_Nature2011}, nuclear spin polarization\cite{Jacques_PRL2009,Ivady_PRB2015}, opto-mechanical spin control\cite{MacQuarrie_natComm2017}, and spin-photon entanglement\cite{Togan2010}, it is expected that understanding the ES of \VB in hBN will be critical to unlocking its potential for quantum technologies.

In this work we perform temperature-dependent continuous-wave (CW) and pulsed optically detected magnetic resonance (ODMR) measurements to manipulate the electronic spin of V$_\text{B}^-$ defects in hBN and reveal the orbital excited-state Hamiltonian. Using confocal microscopy, we excite defect ensembles in multilayer flakes of hBN with a 532 nm laser, and collect the emitted photoluminescence (PL) around $\lambda_{\text{max}} \approx 800$ nm [Supplementary Fig. S1] as a function of applied microwave (MW) frequency to determine the electron spin resonance (ESR) spectrum.  At room-temperature, we measure a zero-field longitudinal splitting $D_\text{es}$ of 2.1 GHz, transverse splitting $E_\text{es}$ of 74 MHz, and a Landé g-factor of 2. Our findings explain the magnetic field-dependent photoluminescence in terms of Zeeman-mediated level anti-crossings in both the ground- and excited-state spin manifolds. Additionally, our temperature-dependent ODMR spectra show that, unlike the NV$^-$ center in diamond\cite{BatalovPRL2009}, the spin-resonance contrast in the ES persists at low temperatures, suggesting that the ES is an orbital singlet. We observe linewidth narrowing and contrast enhancement as the temperature is lowered, which is consistent with previous reports of the temperature-dependent ES lifetime\cite{Liu_BV_tempDep}.

\section{Results}
\begin{figure}[ht]
\centering
\includegraphics[scale=0.35]{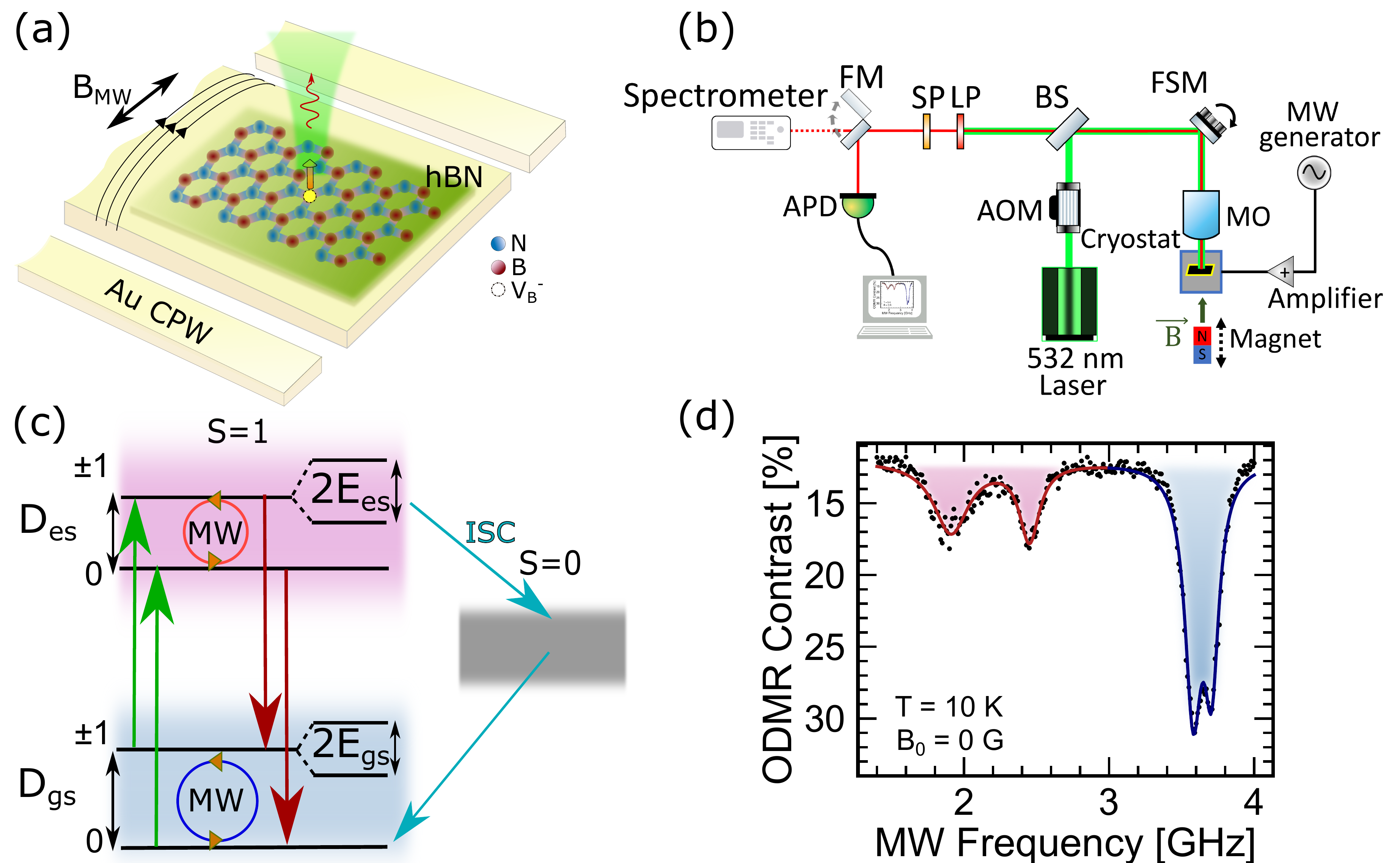}
\caption{\footnotesize
\textbf{Spin-active V$_\text{B}^-$ defects in hBN.} \newline
\textbf{a)} Schematic of the device used to probe the spin transitions of V$_\text{B}^-$ defects in hBN with ODMR. hBN flakes are transferred onto a gold-film coplanar waveguide (CPW) and the generated microwave field $B_\text{MW}$ induces rotations of the defect spin state which is read out via PL. \textbf{b)} Schematic of the experimental setup. AOM = acousto-optic modulator, BS = beam splitter, FSM = fast-steering mirror, MO = microscope objective, LP = long-pass filter, SP = short-pass filter, FM = flip mirror, APD = avalanche photodiode. \textbf{c)} Energy level diagram of the defect orbital states and their spin sublevels, which are split by $D_\text{es}$ and $D_\text{gs}$ in the ES and GS, respectively. A non-radiative ISC to a singlet state is preferred from the $\ket{\pm1}$ spin sublevels of the ES. \textbf{d)} Zero-field ODMR spectrum at T = 10 K, excited with laser power $P_\text{L}\approx 3$ mW and microwave power $P_\text{MW}\approx 160$ mW, showing distinct resonance dips from spin transitions in the ES (red) and GS (blue) at their respective splitting frequencies.}
\label{fig1}
\end{figure}

The V$_\text{B}^-$ defect in hBN has a spin triplet in both the orbital GS and ES \cite{Reimers_PRB_2020,Gottscholl_natureMat}, with the electronic spin oriented out-of-plane with respect to the hexagonal crystal lattice of the hBN. Spin-spin interactions introduce a zero-field splitting between the \mszero and \mspm spin states, where $m_\text{S}$ is the spin quantum number. The room temperature zero-field longitudinal splitting of the GS, $D_\text{gs}$, has been measured to be $\sim$3.5 GHz\cite{Gottscholl_natureMat,Liu_BV_tempDep}, but the splitting in the excited state, $D_\text{es}$, was previously unknown. Optical transitions between orbital states are spin-conserving, with the exception of a non-radiative relaxation mechanism through an intersystem crossing (ISC) to a spin-singlet state that is selectively preferred from $\ket{m_\text{S}=\pm1}$ in the ES [Fig. \ref{fig1}(c)]. This non-radiative relaxation process results in a measurable difference in photoluminescence (PL) intensity, allowing for optical readout of the defect spin state.

To probe the spin transitions of the \VB center, we fabricate devices illustrated by the schematic in Figure \ref{fig1}(a). First, multilayer hBN flakes are mechanically exfoliated onto a silicon substrate with a thermal oxide layer and He$^+$ ion implantation is used to break B-N bonds, creating spin-active V$_\text{B}^-$ defects. We use photolithography to pattern a gold-film coplanar microwave waveguide on a sapphire substrate. Then, hBN flakes with spin-active defects are lifted from the SiO$_2$/Si substrate with a polycarbonate (PC) stamp and transferred directly onto the central conductor of the waveguide so that the generated microwave magnetic field, $\vec{B}_\text{MW}$, is orthogonal to the defect spin axis. ODMR is performed by sweeping the frequency of $B_\text{MW}$ while optically pumping with a 532 nm laser and measuring the photoluminescence intensity ($I_\text{PL}$) with an avalanche photodiode (APD) [Figure \ref{fig1}(b)]. When $B_\text{MW}$ is not resonant with spin transitions, the defect remains spin-polarized in the $\ket{m_\text{S}=0}$ spin state and emits the maximum value of $I_\text{PL}$. However, on resonance, the spin is rotated toward $\ket{m_\text{S}=\pm1}$ and $I_\text{PL}$ is reduced due to the enhancement in non-radiative relaxation through the ISC. From the initial state of $\ket{m_\text{S}=0}$ in the orbital GS, the defect can relax through the ISC if either the spin has been rotated toward the $\ket{m_\text{S}=\pm1}$ spin sublevels in the GS and then optically excited, or if it is first optically excited and then rotated while in the ES. Therefore, under continuous MW driving and optical pumping, spin transitions can be induced in both the GS and ES, which is evident in the low-temperature zero-field ODMR spectrum shown in Figure \ref{fig1}(d). Lorentzian fits to the data are shown in separate colors to distinguish the GS (blue) and ES (red) resonances centered at $D_\text{gs} \pm E_\text{gs}$ and $D_\text{es} \pm E_\text{es}$.

\begin{figure}[ht]
\centering
\includegraphics[scale=0.47]{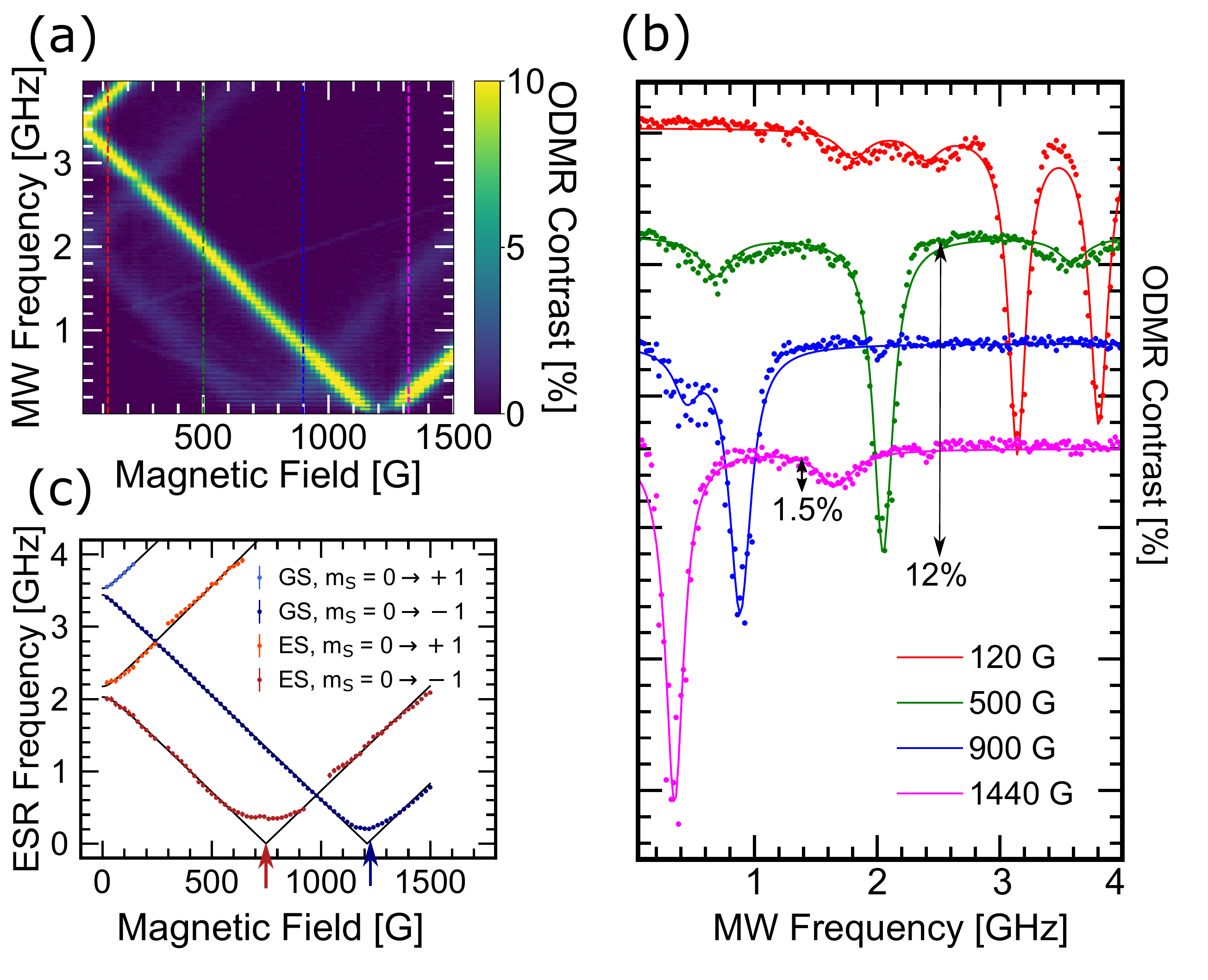}
\caption{\footnotesize
\textbf{Room temperature field dependence of CW ODMR and spin resonance frequencies.} \newline
\textbf{a)} Magnetic field dependence of the CW ODMR contrast at room temperature. \textbf{b)} Individual ODMR traces and their fits at four values of $B_0$ indicated by the vertical colored dashed lines in \textbf{a)}. \textbf{c)} Field dependence of the fitted ODMR peaks corresponding to the electron spin resonance (ESR) frequencies. Fits to equation \ref{esr} are shown in black, from which we extract the values of $D$, $E$, and $g$ in both the GS and ES. The field magnitudes at which the $\ket{0}$ and $\ket{-1}$ spin states cross in energy are marked with red and blue arrows.}
\label{RT_zeeman}
\end{figure}

Due to the Zeeman effect, the degeneracy of the $\ket{m_\text{S}=\pm1}$ states is lifted by an external magnetic field $\vec{B_0}$. Similarly to the ground-state\cite{Gottscholl_natureMat}, the excited-state spin structure can be described by the Hamiltonian
\begin{equation} \label{Hamiltonian}
\hat{H}_\text{es} = \hat{H}_0 + \underbrace{h D_\text{es}(\hat{S_z}^2-\frac{{S}({S}+1)}{3})}_\text{longitudinal splitting}+\underbrace{h E_\text{es}(\hat{S_x}^2-\hat{S_y}^2)}_\text{transverse splitting} +\underbrace{\mu_B g_\text{es}\vec{B_0}\cdot\hat{\vec{S}}}_\text{Zeeman interaction} + \underbrace{\hat{H}_\text{HF}}_\text{hyperfine}
\end{equation}
where $\hat{H}_0$ is the dominant electronic term giving the energy with respect to the GS, $h$ is Planck's constant, $\mu_B$ is the Bohr magneton, $g_\text{es}$ is the ES Landé g-factor, \{$\hat{S}_x,\hat{S}_y,\hat{S}_z$\} are the components of the spin operator $\hat{\vec{S}}$, and S=1 for spin-triplet levels. To study ODMR of Zeeman-split states, we use a permanent magnet aligned perpendicular to the waveguide plane to introduce a static magnetic field. In the limit of a perfectly flat and conformal hBN layer, the field direction will closely coincide with the c-axis of the hBN crystal. From equation \ref{Hamiltonian}, the field magnitude ${B_0}$ shifts the ES spin transition frequencies of the $\ket{m_\text{S}=\pm1}$ states to
\begin{equation} \label{esr}
\nu_\pm = D_{es} \pm \sqrt{{E_{es}}^2 + (\mu_B g_\text{es} B_0/h)^2}
\end{equation}
Figure \ref{RT_zeeman}(a) shows the room-temperature magnetic field-dependent ODMR spectra of the defect ensemble from 0 to 1500 G, at microwave power $P_\text{MW}\approx100$ mW. Line cuts at four values of $B_0$ are shown in Figure \ref{RT_zeeman}(b) along with their fitted curves. The known resonant dips from the $\ket{m_\text{S}=0}\rightarrow\ket{m_\text{S}=\pm1}$ transitions in the GS\cite{Gottscholl_natureMat,Tongcang_plasmon} are present, as well as the additional unreported dips that shift with the applied field along paths parallel to the GS resonance lines, which we attribute to spin transitions in the ES.

To quantify the electron spin resonance (ESR) transition frequencies, we fit each ODMR trace with the appropriate number of Lorentzians, and plot the resonant frequencies as a function of applied field [Fig. \ref{RT_zeeman}(c)]. Note that because the ODMR contrast of the ES resonances are significantly lower than those of the GS, we are not able to accurately fit the ES ESR frequencies around values of $B_0$ where the two overlap, and thus there are some gaps in the ES data. We fit the ESR data with Eq.\ref{esr} to extract the Hamiltonian parameters. Our measurements agree with previously reported\cite{Gottscholl_sensing,Gottscholl_natureMat,Liu_BV_tempDep,Tongcang_plasmon} GS splitting parameters $D_\text{gs}=3.48 \pm 0.02$ GHz and $E_\text{gs}=48.0 \pm 7.1$ MHz, and establish the ES splitting parameters $D_\text{es}= 2.11 \pm 0.03$ GHz and $E_\text{es}= 74 \pm 42$ MHz. We also observe that $g_\text{es}\approx g_\text{gs}\approx 2$, which indicates that the orbital angular momentum does not play a significant role in the ES spin structure. The average ES linewidth over the field sweep at this microwave power is $133 \pm 32$ MHz. These measurements were taken at multiple locations from three different hBN flakes with repeatable results. We do not resolve hyperfine splitting in the ES possibly due to power broadening, the short ES lifetime, and inhomogenous broadening in the defect ensemble. Based on the observed linewidth, we estimate an upper limit of $\sim$100 MHz for the ES hyperfine splitting.

\begin{figure}[ht]
\centering
\includegraphics[scale=0.4]{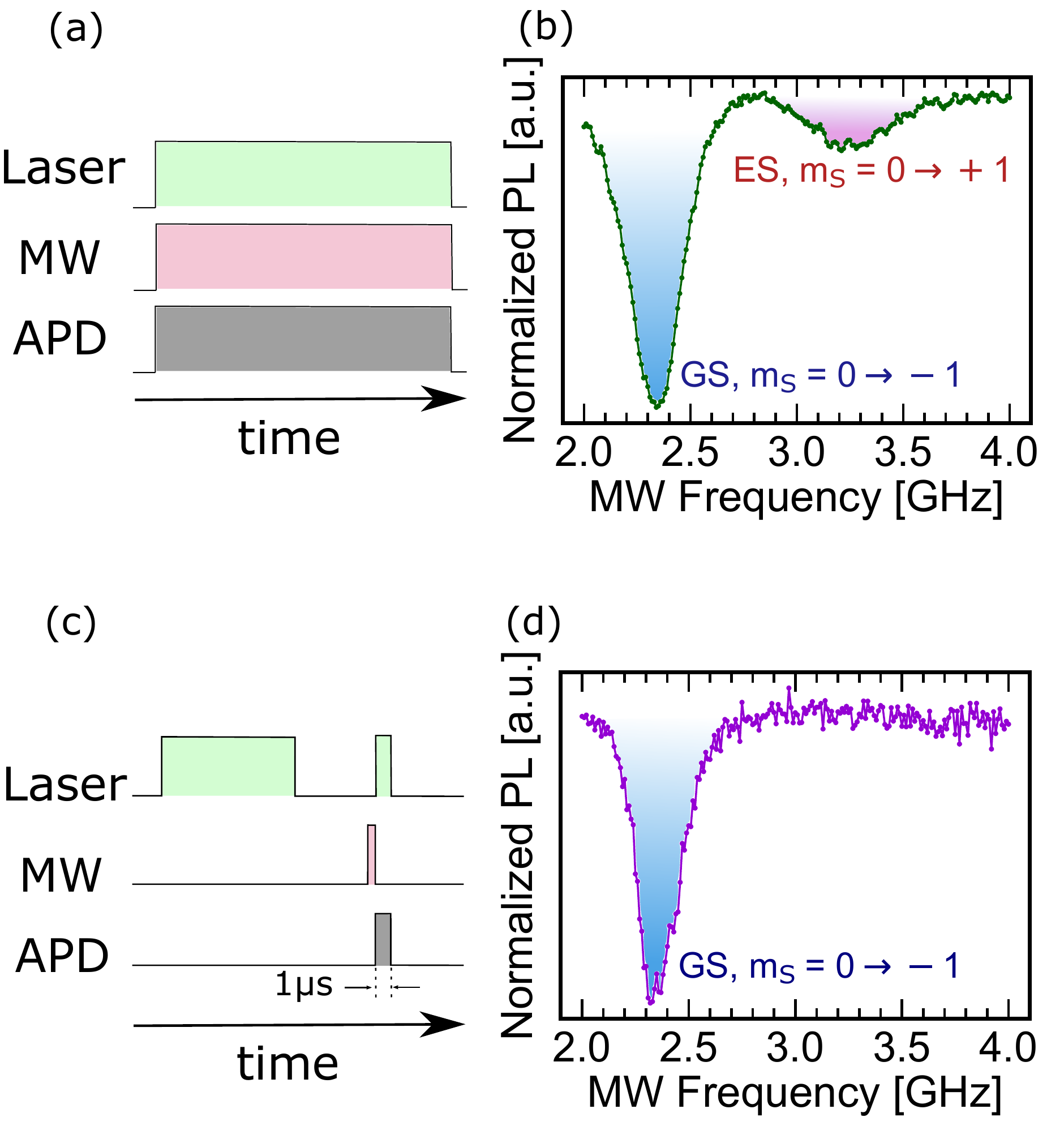}
\caption{\footnotesize
\textbf{Pulsed ODMR at $\boldsymbol{B_0}$ = 400 G.} \newline
\textbf{a)} CW ODMR is performed with the pumping laser, $B_\text{MW}$, and avalanche photodiode (APD) photon counters on at all times. \textbf{b)} The resulting CW ODMR spectrum shows spin transitions in both the GS (blue) and ES (red). \textbf{c)} Pulsed ODMR is performed with $B_\text{MW}$ on only when the laser is off, after the defect is initialized to $\ket{m_\text{S}=0}$ in the GS. As soon as $B_\text{MW}$ is turned off, the laser and photon counters are then turned on to read out the spin state. \textbf{d)} The resulting pulsed ODMR spectrum indicates spin rotation only in the GS, confirming that the additional broader resonance is from an ES transition.}
\label{pulsed_odmr}
\end{figure}

To verify our attribution of the additional ESR frequencies to spin transitions within the orbital excited-state, we implement a pulsed ODMR sequence to compare with CW ODMR results. In the CW measurement, we apply simultaneous microwave and optical excitation [Fig. \ref{pulsed_odmr}(a)], allowing spin rotations to occur both in the GS and in the short time interval spent in the ES, resulting in ODMR contrast at the GS and ES ESR frequencies. In the pulsed measurement, the signals are timed such that the microwave pulse trails the optical excitation pulse by a delay that is much longer than the ES lifetime (1-2 ns) [Fig. \ref{pulsed_odmr}(c)]. In this case, the ES population is negligible when the microwave fields are applied and consequently, the ODMR contrast at the ES ESR frequency disappears if one of the observed spin transitions occurs in the ES. We achieve this by pulsing the laser excitation for 20 µs followed by a waiting period of 5 µs to ensure that the \VB defects have relaxed to the orbital GS with a \mszero spin polarization. While in the dark, we apply a 500 ns burst of $B_\text{MW}$, followed immediately by a 1 µs laser pulse and photon collection for readout. We sweep the frequency of $B_\text{MW}$ from 2 - 4 GHz with a step size of 10 MHz and repeat the measurement $\sim$10$^4$ times at each frequency to build statistics. The results of the CW and the pulsed ODMR measurements at $B_0 =$ 400 G are shown in Fig. \ref{pulsed_odmr}(b) and (d) respectively. In the CW ODMR spectrum, we observe a $\ket{m_\text{S}=0}\rightarrow\ket{m_\text{S}=-1}$ resonance in the GS at $\sim$2.3 GHz and a $\ket{m_\text{S}=0}\rightarrow\ket{m_\text{S}=+1}$ resonance in the ES at $\sim$3.3 GHz. In the pulsed ODMR spectrum, we only see the GS resonance at $\sim$2.3 GHz, confirming that the resonance at $\sim$3.3 GHz indeed belongs to the excited-state manifold\cite{FuchsPRL2008}.

Our observation of a single set of spin-triplet ES resonances [Fig. \ref{RT_zeeman}(a)] is consistent with the ES being either an orbital-singlet or an orbital-doublet where the two branches have identical spin coefficients $D_{es}$, $E_{es}$, and $g_{es}$. Electronic orbitals are determined by the irreducible representations of the point group symmetry of the defect, and theoretical predictions of the \VB center suggest that it has either D$_{3h}$ symmetry with a $^{3}E^{''}$ ES doublet, or C$_{2v}$ symmetry with a $^{3}B_{1}$ ES singlet\cite{Gali2020,Reimers_PRB_2020}. In orbital-doublet systems like the $^{3}E$ ES of the diamond NV$^-$ center, the dynamic Jahn-Teller effect averages the two orbital branches at room temperature\cite{KaiMeiFu_PRL2009}, but at low temperatures the effect is quenched and the ODMR contrast at MW frequency $D_\text{es}$ diminishes\cite{BatalovPRL2009}. To look for signatures of this effect in \VB centers in hBN, we cool our device in an optical cryostat. At T = 10 K, we find that not only does the ES ODMR contrast persist, but also the transverse anisotropy splitting drastically increases from 74 MHz at room temperature to 258 $\pm$ 2 MHz, an effect that is not observed in the GS\cite{Liu_BV_tempDep}. This is possibly due to strain induced by the difference in thermal expansion of the hBN flake and the substrate, which further supports the assumption that the point group is reduced from D$_{3h}$ to C$_{2v}$\cite{Gali2020,Reimers_PRB_2020}. Thus, our low-temperature observations suggest that the defect has C$_{2v}$ symmetry with an orbital-singlet ES.


\begin{figure}[ht]
\centering
\includegraphics[scale=0.6]{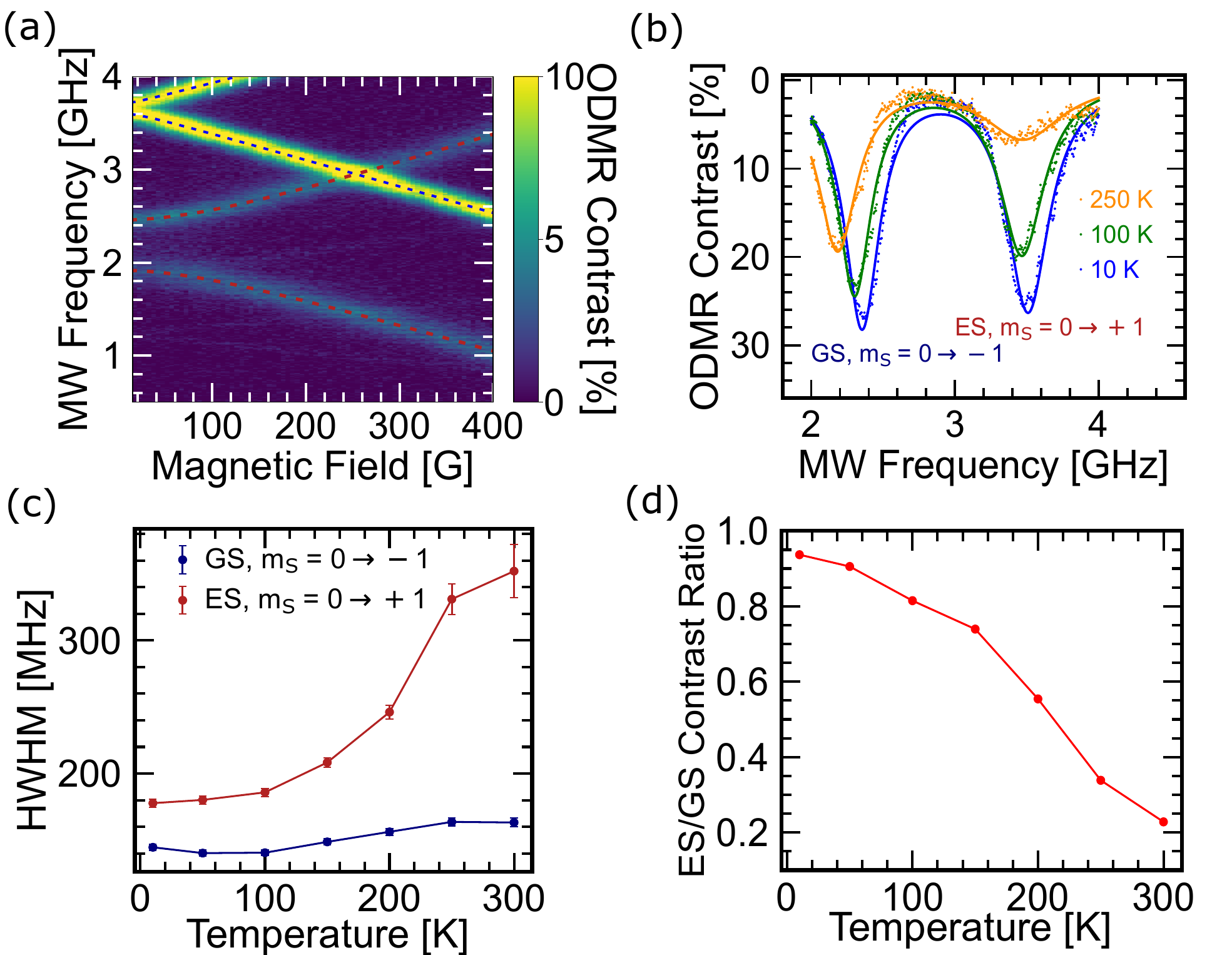}
\caption{\footnotesize
\textbf{Temperature dependence of the ES ESR.} \newline
\textbf{a)} Magnetic field-dependent CW ODMR at T = 10 K and $P_\text{MW}\approx$ 50 mW. ESR fits to equation \ref{esr} have been marked with blue and red dashed lines respectively. \textbf{b)} CW ODMR spectra at T = 250 K (orange), 100 K (green) and 10K (blue) at $B_0=$ 450 G and $P_\text{MW}\approx$ 2.51 W. \textbf{c)} Temperature dependence of the ESR linewidths for both the GS (blue) and ES (red) at $P_\text{MW}\approx$ 2.51 W. \textbf{d)} Relative contrast of the ES ESR with respect to the GS ESR contrast at $P_\text{MW}\approx 2.51$ W.}
\label{tempDep}
\end{figure}

The increased transverse anisotropy splitting is visible in Fig. \ref{tempDep}(a), where we show the CW ODMR at 10 K as we sweep $B_0$ from 0 to 400 G at $P_\text{MW}\approx$ 50 mW. The dashed blue and red curves indicate the fits of the field-dependent ESR frequencies to equation \ref{esr} for the GS and ES, respectively. Next we apply a magnetic field of $B_0=$ 450 G and sweep $P_\text{MW}$ at several temperatures from 10 K to 300 K [Supplementary Fig. S2]. In figure \ref{tempDep}(b), we show the ODMR spectra at three temperatures measured at $P_\text{MW}=$ 2.51 W. From these spectra, it is apparent that the electronic spin is more sensitive to temperature in the ES than in the GS. We find that while the GS ESR linewidth is relatively insensitive to temperature, the ES linewidth decreases with temperature [Fig. \ref{tempDep}(c)]. This suggests an increase in the ES lifetime, which is inversely proportional to the ES ESR linewidth. Most notably, we observe a dramatic enhancement of the ES ODMR contrast at low temperature, approaching the same contrast as the GS ODMR at saturated $P_\text{MW}$ [Fig. \ref{tempDep}(d)]. This is also consistent with a longer ES lifetime at low temperature because a longer average lifetime allows the ES spin to accumulate more rotation toward \mspm  at a given Rabi frequency ($\Omega \propto \sqrt{P_\text{MW}}$). Given that the excited state lifetime is approximately 2 ns at 10 K\cite{Liu_BV_tempDep}, observation of this effect requires $\Omega\approx$ 100 MHz. To confirm our explanation of the temperature-dependent contrast, we crudely model the on-resonance ODMR using $\langle\Delta I_\text{PL}\rangle = \Delta C \sin^2(\frac{\Omega}{2} \tau_\text{es})$, where $\langle\Delta I_\text{PL}\rangle$ is the normalized PL change measured in the defect ensemble, $\Delta C$ is the temperature-independent maximum ESR contrast at saturation ($\sim$25\% at $P_\text{MW} \approx$ 2.51 W), $\Omega$ is the Rabi frequency, and $\tau_\text{es}$ is the temperature-dependent ES lifetime. Using reported ES lifetime values\cite{Liu_BV_tempDep}, our model predicts that the absolute ES ODMR contrast at $T=$ 300 K is $\sim$10\%, which agrees well with our observations [Supplementary Section S3].

Next, we study the influence of magnetic field on the fluorescence of \VB defects at room temperature without microwave excitation. At magnetic fields where the \mszero and \msminus levels cross each other in energy, spin mixing occurs if the defect axis and magnetic field direction are not perfectly aligned [Supplementary Fig. S3]. The off-axis component of $\vec{B_0}$ causes a level anti-crossing (LAC), making the spin eigenstates a mixture of \mszero and \msminus. Because non-radiative orbital relaxation is more efficient for \msminus than for \mszero, mixed spin states have reduced PL intensity, which has been observed in NV$^-$ centers in diamond\cite{Epstein2005}. Using Equation 2 and our measured values of $D_\text{es}$  and $D_\text{gs}$, we expect the excited- and ground-state level anti-crossings (denoted by ELAC and GLAC henceforth) to occur at ${B_0}\approx$ 750 G and 1240 G respectively. These fields are marked with red and blue arrows in figure \ref{RT_zeeman}(c).

\begin{figure}[ht]
\centering
\includegraphics[scale=0.55]{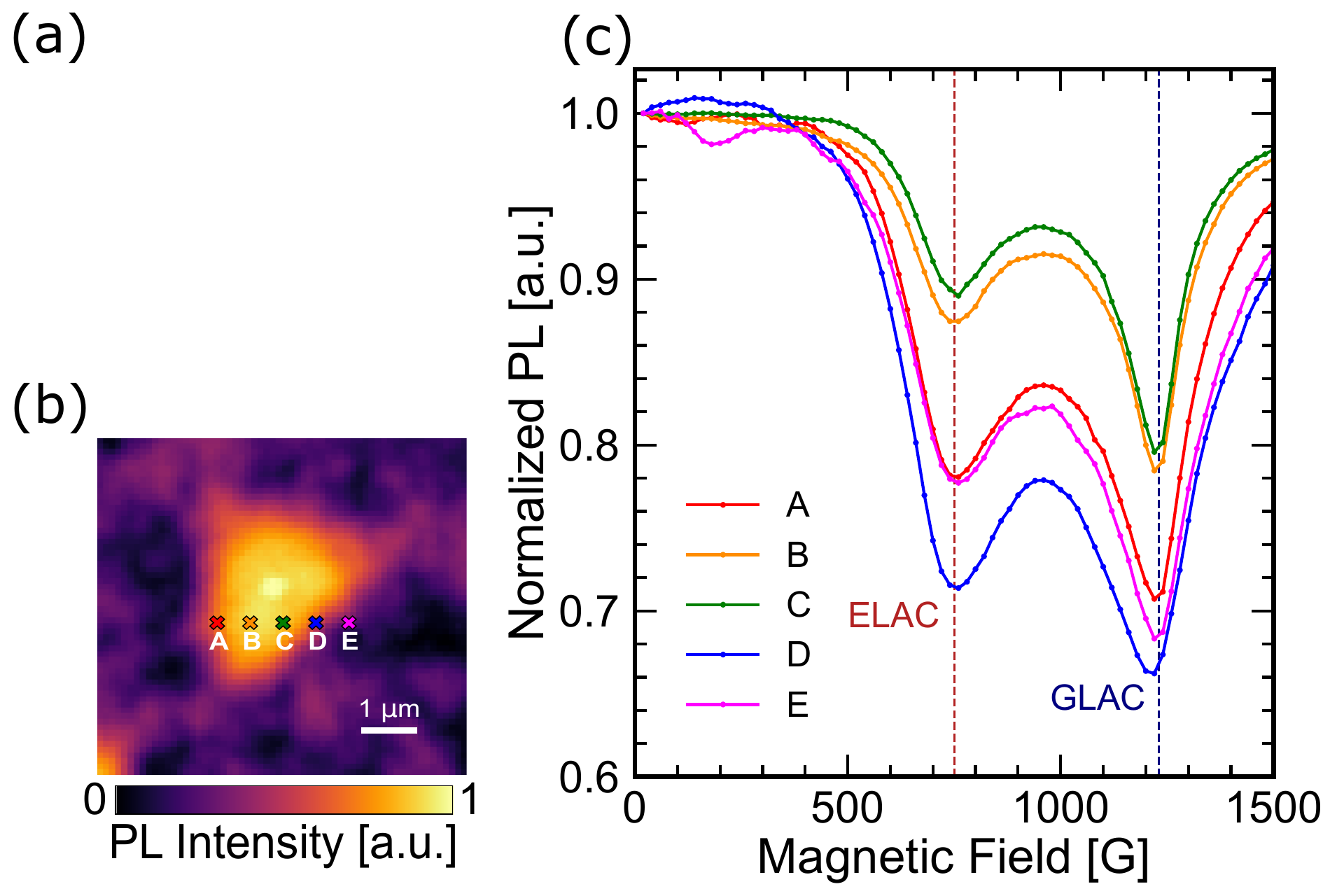}
\caption{\footnotesize
\textbf{Angle-dependent level anti-crossings around a micro-bubble.} \newline 
\textbf{a)} Optical micrograph of an hBN flake on the Au CPW. Micro-bubbles are formed during the transfer process, from which we can study the effect of the magnetic field angle on the level anti-crossings (LACs). \textbf{b)} Spatially resolved PL intensity map of the region indicated by the red dashed box in \textbf{a)}. \textbf{c)} Field-dependent PL measurements collected at the corresponding marked locations in \textbf{b)} show reduction in PL at the ELAC and GLAC field values predicted by our CW ESR results. Variation across locations around the micro-bubble indicate that more spin mixing occurs at greater angles with respect to $\vec{B_0}$. }
\label{LACs}
\end{figure}

In figure \ref{LACs}, we show field-dependent PL measurements in the spatial vicinity of a micro-bubble formed during the transfer process of the hBN flake. Even though the field direction is fixed, we expect that defects at different locations on the bubble will have their defect axes oriented in slightly different directions and thus have different angles with respect to $\vec{B_0}$. Figure \ref{LACs}(a) shows an optical micrograph of one of our devices. A spatially resolved PL intensity map of the region indicated by the red dashed box is shown in Figure \ref{LACs}(b). The bright region is a micro-bubble and the locations of further measurements are indicated by colored markers (labeled A-E). The magnetic field-dependent PL measurements at each of these locations are shown in figure \ref{LACs}(c) and plotted in the corresponding colors. We see two minima that we assign to the ELAC and GLAC. The magnetic field magnitudes for the ELAC and GLAC calculated above agree well with our observations  [Fig. \ref{LACs}(c)]. The PL reduction magnitude varies with location around the bubble, as expected due to the non-uniform topography of the hBN flake\cite{Epstein2005}. These observations further corroborate our understanding of the ES and GS spin-dependent energy level structure.

\section{Discussion}
We have presented a study of the excited-state spin level structure of \VB defects in hBN by performing spin-resonance optical spectroscopy. Our measurements establish the ES room-temperature zero-field longitudinal splitting $D_\text{es}=$ 2.1 GHz, transverse splitting $E_\text{es}=$ 74 MHz, and power-broadened linewidth of 154 MHz. We do not resolve a hyperfine splitting, which we expect is due to the power-broadening that occurs at the large Rabi fields necessary to observe ES ESR. Using pulsed ODMR, we verify that the previously unreported ESR features correspond to spin rotations in the ES, and we observe agreement between magnetic field-dependent PL measurements and a simple model of LACs in the GS and ES. Our temperature-dependent ESR measurements show that the ES ESR contrast around the $D_\text{es}$ frequency not only persists at cryogenic temperatures, but also increases drastically. This and the accompanying linewidth reduction is well-explained by an increase in the ES lifetime. 

The persistence of the ES ESR across the measured temperature range points to an ES orbital-singlet, which naturally lacks a dynamic Jahn-Teller effect, the well-known cause of fast orbital telegraphing in orbital-doublet systems like the NV$^-$ center\cite{KaiMeiFu_PRL2009} that is quenched at low temperature\cite{BatalovPRL2009}. This suggests that the intrinsic D$_{3h}$ symmetry is reduced to C$_{2v}$\cite{Gali2020,Reimers_PRB_2020}. \if A low-temperature photoluminescence excitation (PLE) experiment could confirm this by measuring the optical absorption spectrum to resolve the ES spin sublevels.\fi Our observation of an increase in the transverse anisotropy splitting at low-temperature suggests sensitivity to in-plane strain, which can be explored in future experiments. Additionally, a study of the hyperfine splitting near the ELAC could shed light on the coupling to nuclear spins\cite{FuchsPRL2008}, which could have applications in long-lived spin-based memories\cite{Pfender_NatComm2017} or quantum simulation\cite{Cai_QSimulation_NatPhys2013}. Rapid progress is being made in integrating hBN defects into nano-photonic devices with waveguides\cite{Li2021} and optical cavities\cite{Froch_nanoLett2021,ProsciaMenon2020} to achieve high signal-to-noise ratios for sensing applications. We also envision using \VB defects as quantum sensors\cite{Gottscholl_sensing} for magnetization of layered out-of-plane magnets like CrI$_3$ and CrBr$_3$.

\section{Methods}
\footnotesize
\subsection{hBN device fabrication}
Monocrystalline hBN was tap exfoliated into thin flakes (10 - 100 nm) and transfered onto silicon substrates with 285 nm-thick thermal oxide layers on top. The purpose of the thermal oxide layer is to increase the optical contrast for observing thin hBN flakes. The hBN flakes were irradiated with 2.5 keV He$^+$ ions in a home built ion implanter to create VB- defects. With a flux of $1.6\times10^{11}$ cm$^{-2}$ per second for 10 minutes, the integrated dose density reaches approximately $10^{14}$ cm$^{-2}$. To fabricate the microwave CPW device substrates, first a 3 µm Au film was deposited onto a sapphire substrate using a CVC SC4500 e-beam evaporation tool. Photoresist was spun and a GCA 6300 DSW 5x g-line wafer stepper tool was used to pattern the regions between the ground and central conductor planes. The exposed Au was then removed with a I$_2$/KI Au etchant solution. The hBN flakes with spin active defects were transferred onto the CPW substrates using a stamp consisting of a thin polycarbonate (PC) film mounted on a supporting block of polydimethylsiloxane (PDMS) on a glass microscope slide. The glass slide is placed on a micropositioner in a home-built transfer station and the SiO$_2$ substrate with hBN flakes is held in place on a heated stage with double-sided kapton tape, directly below the polymer stamp. The stamp is lowered until it makes contact with the substrate, upon which the temperature is increased to 80 $^\circ$C, allowing flakes to adhere to the PC as it is lifted off the substrate. Then, the Au CPW substrate is placed on the heated stage and the flake is positioned above the central conductor of the waveguide and lowered until it makes contact with the Au. The heater is then set to 150 $^\circ$C allowing the PC to melt off of the PDMS and adhere to the substrate. The glass slide with PDMS is then lifted and removed, and the sample is gently placed in a chloroform solution to dissolve the PC, leaving behind only the transferred hBN flake. Finally, we fix the device to our sample-holder using a thin layer of wax and use a West Bond 747630E tool to wire-bond the device to the sample-holder, which is then mounted into our cryostat for measurements.

\subsection{Experimental setup for ODMR}
All measurements were performed with a home-built laser-scanning confocal microscope [Fig. \ref{fig1}(b)]. A 532 nm laser (Dragon Lasers) is passed through an acousto-optic modulator (AOM) and focused onto the substrate through a window in a Janis He-flow cryostat using a 50x 0.7 NA Olympus microscope objective. The filtered PL emission is separated from the excitation laser with a beam-splitter followed by a 532 nm notch filter and 637 nm long-pass filter. The PL emission is then focused into a multimode fiber and coupled to an Excelitas avalanche photodiode (APD).
The microwaves are generated by a Stanford Research Systems SG384 signal generator and amplified with either a Mini-Circuits ZVE-3W-83+ high-power amplifier or Mini-Circuits ZVA-183-S+ wideband amplifier before being transmitted into the cryostat. For CW ODMR measurements, a Stanford Research Systems DG645 digital delay generator (DDG) sends pulses to modulate the SG384 output as well as a Mini-Circuits ZYSWA-2-50DR switch that directs the APD counts to two channels on a National Instruments DAQ for normalization of the PL during ODMR frequency sweeps. For pulsed ODMR measurements, the DDG sends the pulse scheme described in the text and illustrated in figure \ref{pulsed_odmr}(c) to the AOM, SG384, and RF switch.

\normalsize
\section{Funding Acknowledgments}
\footnotesize
This work was supported by the Cornell Center for Materials Research (DMR-1719875), an NSF MRSEC, and by the AFOSR (FA9550-18-1-0480). J.L acknowledges support from the NSF (ECCS-1839196). A.N.V acknowledges support from AFOSR (FA9550-19-1-0074). T.L. and X.G. thank the support from the DARPA NLM program and the DARPA QUEST program. Device fabrication was performed at the Cornell Nanoscale Facility, a member of the National Nanotechnology Coordinated Infrastructure (NNCI), which is supported by the NSF (NNCI-2025233).

\normalsize
\section{Author Contributions}
\footnotesize
N.M, A.M., A.N.V., and G.D.F. conceived and designed the experiments. X.G. exfoliated hBN flakes and performed ion implantation to create \VB defects. J.L. and N.M. fabricated the devices. N.M., B.A.M., and A.M. developed the software and apparatus for data acquisition. N.M. and A.M. performed all measurements. N.M., A.M., A.N.V, and G.D.F. analysed the data. A.N.V. and G.D.F. supervised the project. N.M. and A.M. wrote the manuscript with input from X.G., B.A.M., T.L., A.N.V., and G.D.F.

\bibliographystyle{unsrt}  



\newpage

\normalsize

\section*{\LARGE{Supplementary Information}}

\section*{S1: Photoluminescence spectra of \VB ensembles}

As shown in figure 1(b) in the main text, we excite the \VB defect ensemble with a 532 nm laser in a confocal microscope setup. The emitted photoluminescence (PL) spectrum is shown in figure S1. We note that there is no discernible sharp zero-phonon line (ZPL), but we do observe a broad fluorescence band centered around 800 nm which likely represents phonon-mediated emission. For ODMR measurements, we isolate the emission from the spin-active defects by filtering the PL with a 670 nm longpass filter. 
\begin{figure}[ht]
\centering
\includegraphics[scale=0.6]{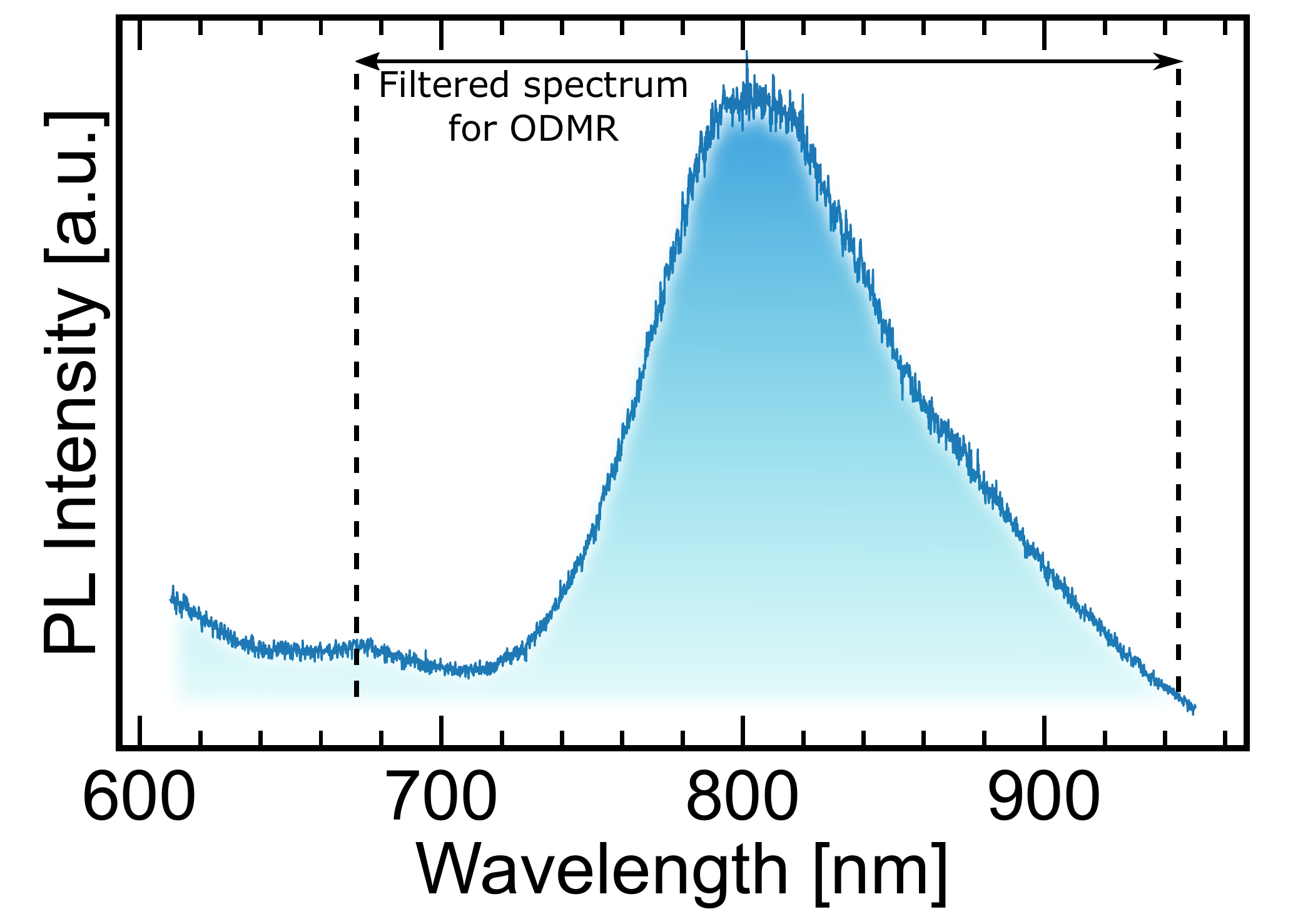}
\caption*{\textbf{Figure S1. Room-temperature optical spectrum of \VB defect emission}
\newline}
\label{SI_spectrum}
\end{figure}

\section*{S2: Microwave power-dependent ODMR}
\begin{figure}[h]
\centering
\includegraphics[scale=0.8]{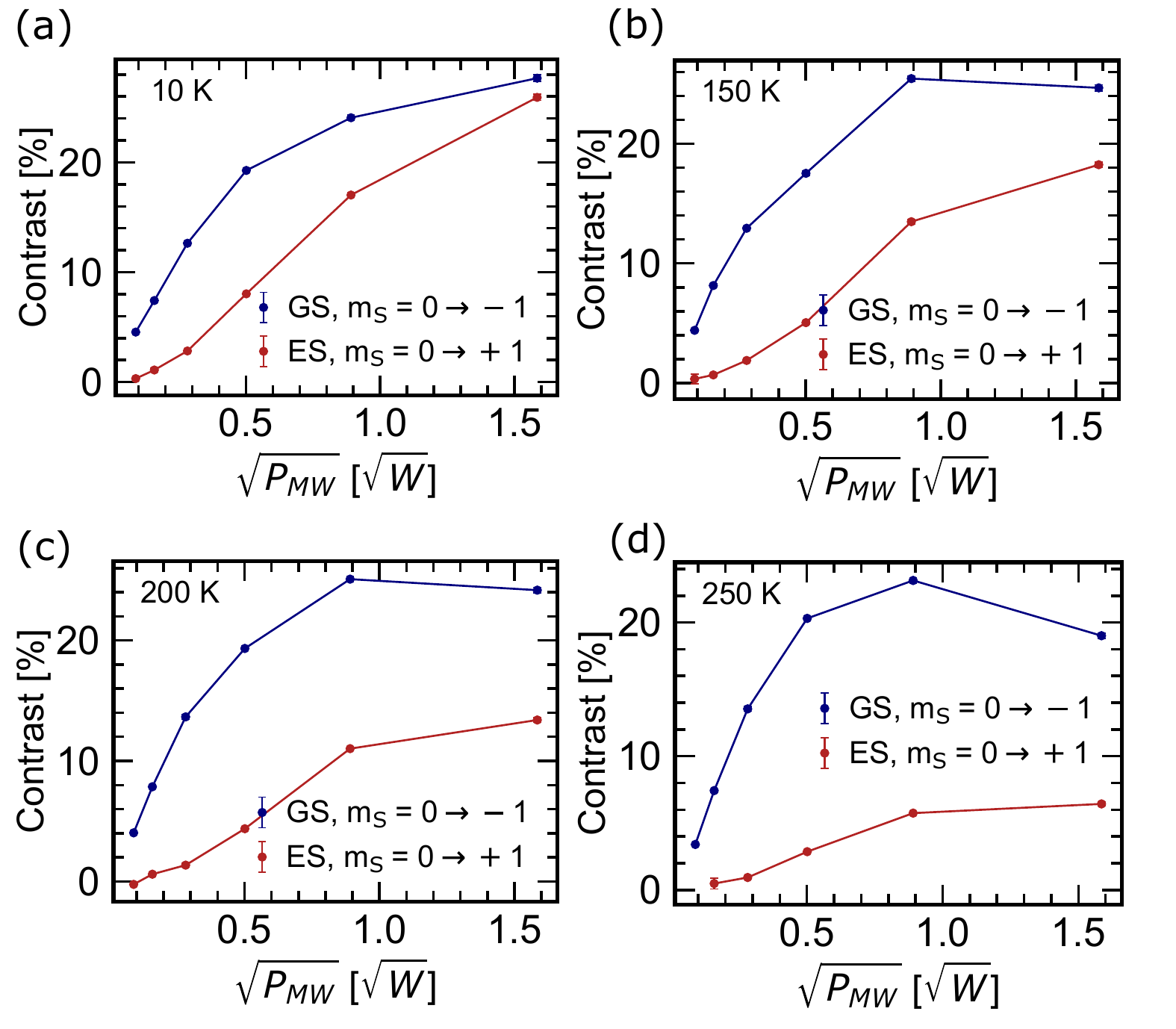}
\caption*{
\textbf{Figure S2. Microwave power-dependent ODMR} at temperatures \textbf{(a)} 10 K, \textbf{(b)} 150 K, \textbf{(c)} 200 K, and \textbf{(d)} 250 K.}
\label{SI_powerSweeps}
\end{figure}
Microwave power ($P_\text{MW}$) sweeps were performed at several temperatures from 10 K to 300 K, with an applied magnetic field of $B_0=$ 450 G. In figure S2, we show the the power-dependence of the ODMR contrast for both the GS (blue) and ES (red) spin transitions. The ES lifetime increases as the temperature is lowered\cite{Liu_BV_tempDep}, allowing the ES contrast at a given Rabi frequency to increase. As a result, the power saturation curve of the ES approaches the curve of the GS as the temperature is lowered. At 10 K, the ES/GS contrast ratio is close to 1 at the highest microwave power in the sweep ($P_\text{MW} \approx$ 2.51 W) [also see Fig. 4(d) in the main text].

\section*{S3: Temperature-dependent ES lifetime model}
The observed temperature dependence of the ES ODMR contrast can be explained by considering the temperature dependence of the ES lifetime. As an example, we consider T = 10~K and 200~K where reported values are $\sim$1.95~ns and $\sim$1.6~ns respectively\cite{Liu_BV_tempDep}. Because the lifetimes are short, it is reasonable to assume that the Rabi-rotation toward the $\ket{\pm 1}$ states is small. In this regime, we model the on-resonance temperature-dependent ODMR contrast, C as:
\begin{equation}
     C (T)= \Delta C \text{sin}^2\left(\frac{\Omega \tau_{es}(T)}{2}\right)\approx \Delta C\frac{\Omega^2 \tau_{es}(T)^2}{4} 
\end{equation}
where $\Delta C$ is the value of the ODMR contrast with full population transfer to the $\ket{\pm 1}$ states, $\Omega$ is the Rabi-frequency, and $\tau_{es}(T)$ is the temperature-dependent lifetime. From figure S2(a), we note the value of $ C(10~K)\sim0.25$. Assuming that $\Omega$ is independent of temperature, we estimate that $ C(200~K)=C(10~K)\left(\frac{\tau_{es}(200~K)}{\tau_{es}(10~K)}\right)^2\approx 0.25\frac{1.6^2}{1.95^2}=0.16$  . Our observed value [Fig. S2(c)] is 0.14 which agrees closely with the calculations.

\section*{S4: Level anti-crossing in the orbital excited and ground states}

\begin{figure}[ht]
\centering
\includegraphics[scale=0.85]{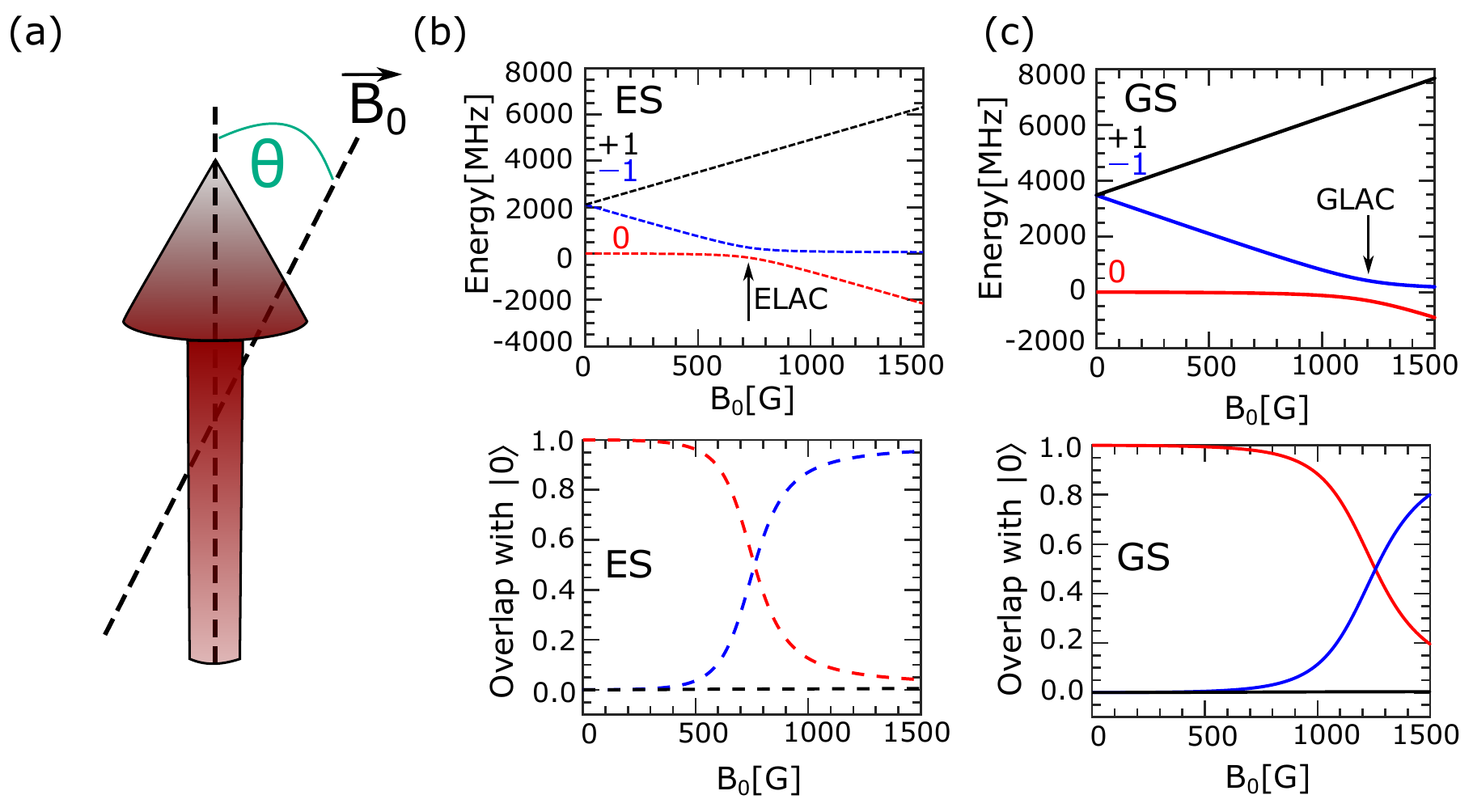}
\caption*{
\textbf{Figure S3. Model of level anti-crossings (LACs) and spin-mixing}
\newline
\textbf{a)} Magnetic field $\Vec{B_0}$ applied at an angle $\theta$ with respect to the defect axis shown along the vertical direction. \textbf{b-c)} Level anti-crossing (top) and spin mixing (bottom) at $\theta=6^\circ$ for the orbital excited state (b) and ground state (c).}
\label{SI_LAC}
\end{figure}

The essence of level anti-crossing (LAC) and spin-mixing in the orbital excited and ground states can be captured by studying the Hamiltonian:
\begin{equation}
\hat{H}=\mu_B g\Vec{B_0}\cdot\Vec{\hat{S}}+hD\hat{S_z^2}=\mu_B gB_{0z}\hat{S_z}  cos \theta+\mu_B g B_{0x}\hat{S_x} sin \theta+hD\hat{S_z^2}
\end{equation}
Here we have assumed that the g-factor of the defect is isotropic and have neglected the transverse splitting parameter $E\ll D$ at room temperature. Without loss of generality, we have also assumed that the applied field $\Vec{B_0}$ lies in the x-z plane at an angle $\theta$ [Fig. S3(a)] with the defect axis along z. We show the spin eigen-energies and overlap of the eigenvectors with the $\ket{0}$ Zeeman sub-level in the orbital excited and ground states in figures S3(b) and S3(c)  respectively. The energy eigenvalue plots show that the LACs occur at $\sim$750 G and $\sim$1240G in the excited and ground states, which match very well with our observations. At the LAC, the eigenvectors overlap with the $\ket{0}$ and $\ket{-1}$ equally, confirming spin-mixing. 

In the main text, we have shown that the magnitude of the photo-luminescence (PL) intensity drops at the LAC fields. We attribute this effect to mixing of $\ket{0}$ with $\ket{-1}$ resulting in an increase in non-radiative relaxation. The magnetic field-dependent PL is governed by the steady-state populations of the $\ket{0}$ and $\ket{-1}$ levels which are dependent on the competing phenomena of spin-polarization from optical pumping and spin-mixing\cite{Epstein2005}. A Lindbladian approach with appropriate optical pumping, inter-system crossing, and spin-relaxation rates is needed to fully explain our observations which is beyond the scope of this work.     



\end{document}